# Design Guidelines For a Thermoacoustic Refrigerator

**Ram C. Dhuley, M.D. Atrey**

*Department of Mechanical Engineering*
*Indian Institute of Technology Bombay, Powai*
*Mumbai-400 076*

*Development of refrigerators based on Thermoacoustic technology is a novel solution to the present day need of cooling without causing environmental hazards. With added advantages of absence of moving parts and circulating refrigerants, these devices can attain very low temperatures maintaining a compact size. The present theoretical work is based on theory of linear thermoacoustics[1]. Under the short stack and invicid assumptions, an algorithm for design of a standing wave thermoacoustic refrigerator, with main focus on the stack, is described. A stack is designed for a given cooling requirement of the refrigerator and certain chosen operation parameters.*

**Nomenclature**

| | | | |
|---|---|---|---|
| p | pressure (N m$^{-2}$) | $\dot{Q}_c$ | cooling power (W) |
| T | temperature (K) | $\dot{W}$ | acoustic power (W) |
| f | frequency (Hz) | COP | coefficient of performance |
| ρ | density (kg m$^{-3}$) | L | length of resonator (m) |
| a | sound speed (m s$^{-1}$) | $L_s$ | length of stack (m) |
| λ | wavelength (m) | A | cross section area of stack (m$^2$) |
| k | wave number (m$^{-1}$) | l | half plate thickness (m) |
| ω | angular frequency (rad s$^{-1}$) | $y_0$ | half plate spacing (m) |
| γ | ratio of specific heats | Π | plate perimeter (m) |
| $δ_k$ | thermal penetration depth (m) | $x_s$ | stack centre position (m) |
| β | thermal expansion coefficient (k$^{-1}$) | | |
| K | thermal conductivity (W m$^{-1}$ k$^{-1}$) | *Subscripts* | |
| $C_p$ | isobaric specific heat (J kg$^{-1}$ k$^{-1}$) | m | mean |
| Γ | normalized temperature gradient | a | amplitude |
| | | 1 | local amplitude |
| s | specific entropy (J kg$^{-1}$ k$^{-1}$) | p | per plate |

## INTRODUCTION

*'Thermoacoustics'* is the interaction between heat and sound. Of the several applications[1,2] of thermoacoustics in energy conversion, the thermoacoustic refrigerator is a well known device. A thermoacoustic refrigerator uses the energy of sound or pressure waves to bring about refrigeration. The main components of a thermoacoustic refrigerator are the resonator, the stack, the acoustic driver and the heat exchangers as shown below in *Figure 1*.

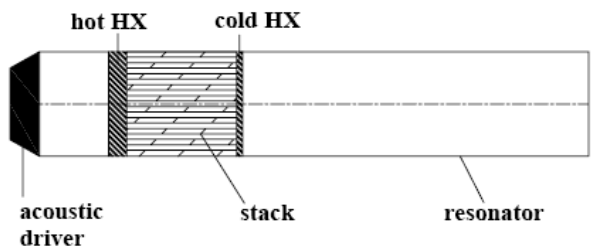

Figure 1  Schematic of thermoacoustic refrigerator

The standing pressure wave generated by acoustic driver in the resonator causes to-and-fro motion of the gas from one end of the stack to the other thereby providing a means of transporting heat. The gas near the pressure node gets cooled due to rarefaction and picks up heat from the stack making one end of stack cold. This gas moves towards pressure antinode and gets heated up due to compression. It loses heat to the stack and makes the other end of stack hot. Thus, a temperature gradient is set up along the stack length. With heat exchangers at the two ends of the stack, this temperature gradient can be used to produce refrigeration.

A detailed description of thermoacoustic cooling cycle has been given by Arnott *et. al*[3]. Braun *et. al*[4] have developed a design optimization program based on the thermoacoustic simulation program known as DELTAE developed by Swift[1]. A design optimization procedure based on normalization of parameters has been given by Wetzel *et. al*[5] and Tijani *et. al*[6]

## OPERATING AND DESIGN PARAMETERS[6]

The aim of the present analysis is to develop a design procedure for a thermoacoustic refrigerator and to determine the design parameters based on certain given operating parameters. Various operating and design parameters are given in *Table 1*.

The properties of working gas like density, thermal conductivity, ratio of isobaric to isochoric specific heats and speed of sound in gas play an important part in this analysis. The working gas should have a low boiling point, high sound speed and should be inert to the components of the refrigerator.

Table 1  Operating and Design parameters

| Operating parameters | Design Parameters |
|---|---|
| mean pressure, $p_m$ | resonator length, L |
| mean temperature, $T_m$ | stack length, $L_s$ |
| frequency, f | stack cross section area, A |
| pressure amplitude, $p_a$ | plate thickness, 2l |
|  | plate spacing, $2y_0$ |
|  | stack centre position, $x_s$ |

## ASSUMPTIONS[1,6]

Following are the assumptions made in the present analysis

1. The length of the stack is much smaller as compared to the wavelength of the standing wave $L_s<<\lambda$ (Short stack approximation). It can be assumed that the local pressure and velocity amplitude of the oscillating gas molecules is more or less same over the entire length of the stack.

2. The flow of the gas in the stack is assumed to be invicid. Friction at the inner walls of the resonator is also neglected.

3. The temperature difference across the stack ends is assumed to be small as compared to the mean temperature $\Delta T_m \ll T_m$. It can be assumed that the thermo-physical properties of the gas do not vary significantly over the stack length and hence can be assumed constant.

4. The whole analysis is carried out at steady state operation of the refrigerator. The mean temperature of the gas is $T_m$ and the temperature gradient across the stack remains constant with time.

5. Conductivity of plate material is neglected.

**GOVERNING EQUATIONS[1]**

The equation governing the entire analysis is the general heat transfer equation expressed in terms of entropy transport. Neglecting viscosity of the gas, it can be written as

$$\rho T \left( \frac{\partial s}{\partial t} + \bar{v} \cdot \nabla s \right) = \nabla \cdot (K \nabla T) \quad (1)$$

where s is specific entropy. The equation states that the rate of change of amount of heat at a certain point depends on convective flow of heat by means of entropy and conduction of heat. Various other expressions[1] relevant to present analysis such as those for the cooling power and the acoustic power can be derived from eq(1).

Referring to *Figure 2*, the acoustic driver produces a region of maximum pressure variation at x=0 i.e a pressure antinode and velocity node. Subsequently, following equations can be written down for local pressure and velocity amplitudes in the resonator :-

$$p_1 = p_a \cos(kx) \quad (2)$$

$$u_1 = \frac{p_a}{\rho_m a} \sin(kx) \quad (3)$$

where k=2π/λ. On account of the porosity 'B' of the stack, eqn.3 gets modified to,

$$u_1 = \frac{p_a}{\rho_m a B} \sin(kx) \quad (4)$$

The critical temperature gradient is an important parameter governing the refrigeration action of a thermoacoustic device. It is given by,

$$\nabla T_{crit} = \frac{(\gamma - 1)kB}{\beta} \cot(kx) \quad (5)$$

where β is the thermal expansion coefficient of the gas. For ideal gas, it is equal to the inverse of the absolute temperature. Of utmost concern in the design procedure are the cooling power produced by the stack, the work input needed to produce this cooling power and thus the COP. These are given by the following expressions :-

$$\dot{Q}_p = -\frac{\delta_k \Pi T_m \beta p_1 u_1 (\Gamma - 1)}{4} \quad (6)$$

$$\dot{W}_p = \frac{\delta_k \Pi L_s T_m \beta^2 \omega (p_1)^2 (\Gamma - 1)}{4 \rho_m C_p} \quad (7)$$

$$COP = \frac{T_m \beta a}{B L_s (\gamma - 1) \omega} \tan(kx) \quad (8)$$

where, $\delta_k$ is the thermal penetration depth of the gas and is an important parameter governing the plate spacing of the stack. It is defined as the distance through which heat diffuses in the gas in time 1/ω and is given by

$$\delta_k = \sqrt{\frac{2K}{\rho_m C_p \omega}} \quad (9)$$

Π is the perimeter of the cross section of the plate. Assuming the thickness of the plate to be very small, the perimeter is equal to twice the plate width ($\Pi=2w$). $\Gamma$ is the ratio of actual temperature gradient to the critical temperature gradient of the stack.

**DESIGN PROCEDURE**

This section describes the design procedure of a refrigerator producing a cooling power of 4 W and a cold end temperature of 210 K. The operating parameters chosen are shown in Table 2[6].

Table 2  Operating and working gas parameters

| Operating parameters | Helium properties |
|---|---|
| $p_m$=10 bar | $\rho_m$=1.9244 kg/m$^3$ |
| $T_m$=250 K | a=935 m/s |
| f=400 Hz | k=0.14 W/mK |
| $p_a$=0.2 bar | $\gamma$=1.67 |
| $T_h$=283 K | |

**Determining the design parameters**

*Resonator length*

The smallest possible length of the resonator which will produce a standing wave is equal to quarter of the wavelength. A standing wave can be generated by keeping one end of resonator closed (driver end) and simulating the other end as 'open'. This can be done by attaching a sufficiently large buffer[6,7] of gas at $T_m$ at this end. For this most fundamental case we get L=58.8 cm.

*Plate spacing*

The thermal penetration depth of the gas at the given frequency is found out to be 0.106 mm. Figure 3 shows variation of heat flux with distance from the plate. It can be seen that almost all the transport of heat takes place in region within $\delta_k$ from the plate and hence the plate spacing is chosen to be 2$\delta_k$[8]. Hence,

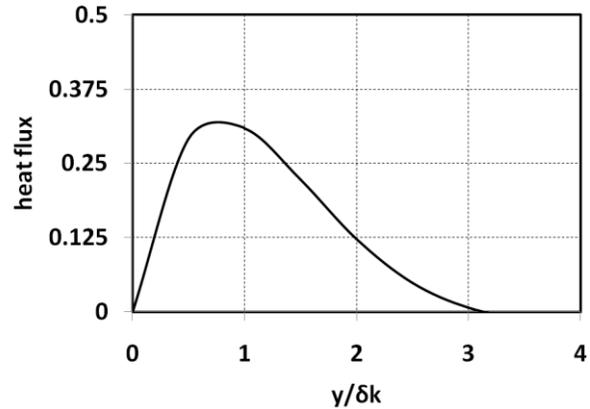

Figure 3  Heat transport *vs.* distance from plate.

*Plate thickness*

The plate thickness can be found out from the expression

$$2l = 2y_0(\frac{1}{B}-1) \qquad (10)$$

where B is the porosity of the stack. The porosity is defined as the ratio of open area in the cross section to the total cross section area of the stack. The porosity is chosen such as not to disturb the acoustic standing wave significantly and is normally taken in the range 0.7-0.8[5,6]. Choosing B=0.75, we get 2l=0.07 mm.

*Stack length, centre position and area of cross section*

A graphical approach has been implemented to determine the stack length, centre position and the cross section area.
    It is assumed that the stack is made of 'n' parallel plates over one another and the cross section of the stack is a square of width 'w'. In this case, the perimeter of the plate cross section becomes 4nl+4(n-1)$y_0$. Knowing the fact that total cooling power produced by

the stack is 'n' times that produced by a single plate, it can be written

$$\dot{Q}_c = -n[n(l+y_0)-y_0] \times \delta_k p_1 u_1 [\frac{\Delta T_m}{L_s \nabla T_{crit}} - 1] \quad (11)$$

Putting the known quantities in eq(11), an equation in three variables x, $L_s$ and n is obtained,

$$0.282n^2 - 0.212n + \frac{509.54}{\sin(2kx)(0.221\frac{\tan(kx)}{L_s}-1)} = 0 \quad (12)$$

Similarly, from eq(8) we get,

$$COP = 0.75\frac{\tan(kx)}{L_s} \quad (13)$$

COP and n are plotted with $L_s$ for different positions of stack centre in the resonator. Theoretically, every point on such a curve will give us a stack producing a cooling power of 4 W and a cold end temperature 210 K. *Figure 4* and *Figure 5* show variation of COP and n respectively with $L_s$ at different stack centre locations.

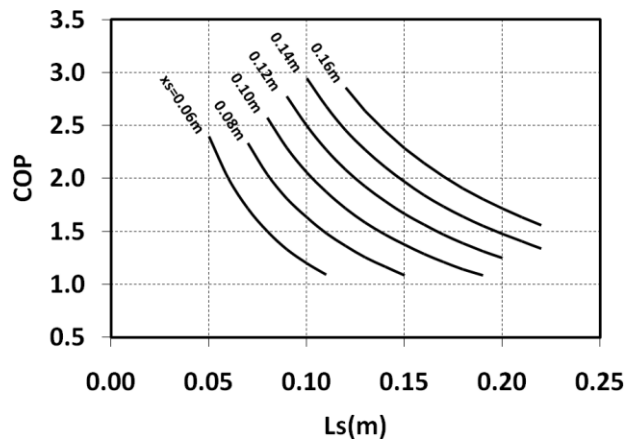

Figure 4  Variation of COP with stack length

For instance, if $L_s$=0.1 m is chosen at $x_s$=0.1 m, a stack with 95 parallel plates is obtained. The cross section area of this stack comes out to be 7.06 cm². This stack will produce a cooling power of 4 W at COP close to 2.

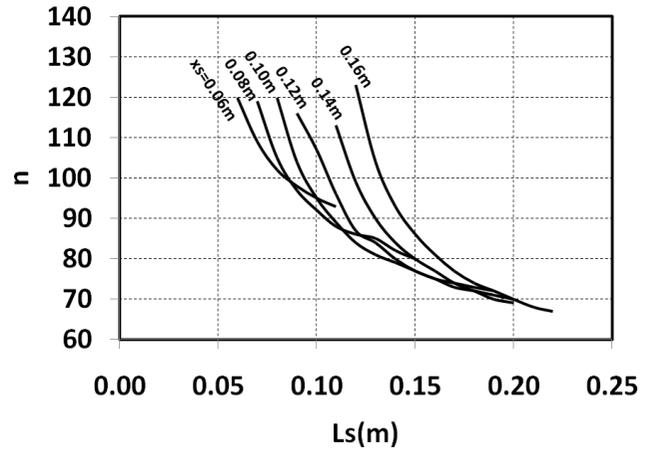

Figure 5  Variation of number of plates with stack length.

## RESULTS AND DISCUSSIONS

The design parameters obtained from the analysis are shown in Table 3

Table 3  Results

| Parameter | Value |
|---|---|
| L | 588 mm |
| $L_s$ | 100 mm |
| A | 706 mm² |
| 2l | 0.07 mm |
| $2y_0$ | 0.212 mm |
| $x_s$ | 100 mm |

The stack length ($L_s$=10 cm) is very small as compared to the wavelength of the acoustic field (λ=235.2 cm). Hence, the assumption of short stack is perfectly valid. At the stack centre location ($x_s$=10 cm) the critical temperature gradient is 1235.13 K/m. The actual temperature gradient across the stack ends is 750 K/m. Clearly, Γ=0.607<1 which is primary necessity[1] for the device to function as a refrigerator, is met.

Another perfectly valid choice of stack producing the same cooling power at same cold end temperature is $L_s$=8.5 cm and $x_s$=8cm. In this case, the stack will consists of 102 plates and a cross section of 8.2 cm$^2$. The corresponding COP is 1.9. Tijani *et al.*[6] obtained a stack with cross section 11.8 cm$^2$ and COP of 1.5 with same stack length and set of operation parameters. The difference in the cross section can be accounted for the authors in [6] had taken viscosity into account and hence needed more number of stack plates to produce 4 W power.

Referring to *Figure 4*, high values of COP (~3) are expected from the analysis. But, it should be noted that this is COP of the stack and not of the refrigerator. Even when the viscosity is neglected, more power has to be input to account for losses at heat exchangers, driver, *etc.* (though not a part of this analysis).

**CONCLUSIONS**

A simple design procedure for a standing wave thermoacoustic refrigerator has been described. With a choice of operating parameters and helium as working gas, graphical approach has been used to determine the geometrical parameters of the stack. A refrigerator producing a cooling power of 4W at cold end temperature of 210K at steady state is designed theoretically using this design procedure.

**References**

1. Swift G.W., Thermoacoustic engines*, J Acoust Soc. Am*, **84**, (1998), pp 1146-1180

2. Swift G.W., Thermoacoustics- A unifying perspective of some engines and refrigerators, *Acoustical Society of America Publications (2002).*

3. Arnott W, Raspet R, Bass H.E, Thermoacoustic Engines, *Ultrasonics Symposium,* (1991)*, pp 995-1003.*

4. Paek I.,Braun J.E., Mongeau L., Evaluation of standing-wave thermoacoustic cycles for cooling applications, *International Journal of Refrigeration*, **32**, (2007), pp 1059-1071.

5. Wetzel M., Herman C., Design Optimization of thermoacoustic refrigerators, *International Journal of Refrigeration*, **20**, (1997), pp 3-21.

6. Tijani M.E.H., Zeegers J.C.H., deWaele A.T.A.M.., Design of thermoacoustic refrigerators*, Cryognenics*, **42**, (2002), pp 49-57

7. Tijani M.E.H., Zeegers J.C.H., deWaele A.T.A.M.., Construction and working of a thermoacoustic refrigerator*, Cryognenics*, **42**, (2002), pp 59-66

8. Tijani M.E.H., Zeegers J.C.H., deWaele A.T.A.M., Optimal stack spacing for thermoacoustic Refrigeration, *J Acoust Soc. Am*, **112**(1), (2002), pp 128-133